\documentclass[aps,amsmath,amssymb,nofootinbib,showpacs,superscriptaddress,showkeys]{revtex4-2}

\usepackage[english]{babel}
\usepackage{latexsym}
\usepackage{graphics}
\usepackage{graphicx}
\usepackage{epsfig}
\usepackage{color}
\usepackage{bm}
\usepackage{amsmath}
\usepackage{amssymb}
\usepackage{amsthm}
\usepackage{dcolumn}
\usepackage{bm}
\usepackage{float}
\usepackage{xr-hyper}
\usepackage{hyperref}
\usepackage{color}
\usepackage{epstopdf}
\usepackage{cleveref}
\usepackage[autostyle]{csquotes}
\usepackage{soul,color}
\usepackage[svgnames]{xcolor}
\hypersetup{hidelinks,colorlinks=true,allcolors=Blue}
\theoremstyle{remark}

\begin{document}

\renewcommand{\thefootnote}{\fnsymbol{footnote}}

\preprint{APS/123-QED}

\title{A machine learning potential-based generative algorithm for on-lattice crystal structure prediction}
\author{Vadim Sotskov*}
\affiliation{Skolkovo Institute of Science and Technology, Skolkovo Innovation Center, Bolshoi Blv. 30, Building 1, Moscow 121205, Russia}
\author{Evgeny V. Podryabinkin}
\affiliation{Skolkovo Institute of Science and Technology, Skolkovo Innovation Center, Bolshoi Blv. 30, Building 1, Moscow 121205, Russia}
\author{Alexander V. Shapeev}
\affiliation{Skolkovo Institute of Science and Technology, Skolkovo Innovation Center, Bolshoi Blv. 30, Building 1, Moscow 121205, Russia}

\footnotetext{Vadim.Sotskov@skoltech.ru}

\date{\today}
\begin{abstract}
We propose a method for crystal structure prediction based on a new structure generation algorithm and on-lattice machine learning interatomic potentials. Our algorithm generates the atomic configurations assigning atomic species to sites of the given lattice, and uses cluster expansion or low-rank potential to evaluate their energy.
We demonstrate two benefits of such approach.
First, our structure generation algorithm offers a ``smart'' configurational space sampling, targeting low-energy structures which significantly reduces computational costs.
Second, the application of machine learning interatomic potentials significantly reduces the number of DFT calculations.
We discuss how our algorithm resembles the latent diffusion models for image generation.
We demonstrate the efficiency of our method by constructing the convex hull of Nb-Mo-Ta-W system, including binary and ternary Nb-W and Mo-Ta-W subsystems. We found new binary, ternary, and quaternary stable structures that are not reported in the AFLOW database which we choose as our baseline.
Due to the computational efficiency of our method we anticipate that it can pave the way towards efficient high-throughput discovery of multicomponent materials.

\end{abstract}

\maketitle

\section{Introduction}
Development of computational technologies in materials science past two decades resulted in paradigm shift in discovery of new materials. Now search for new materials is done by computers, while experimental synthesis takes place at the final stage and only for the most promising and theoretically proven candidates. This approach stimulated the development of large materials databases with the results of computational search like NOMAD \cite{nomad}, OQMD \cite{oqmd}, AFLOW \cite{aflow}, and Materials Project \cite{matproj}, which in their turn enabled the application of data-driven approaches in materials discovery \cite{datamat1,datamat2,datamat3,datamat4}.
These approaches can be divided in two categories: the first provides stoichiometry and optionally some properties of the promising candidates for new materials, while the second include methods predicting the crystal structure. The methods of the second group are considered to be more reliable and provide a possibility of additional study of the thermodynamic stability and some properties.  
\\
Crystal Structure Prediction (CSP) can be considered as a problem of finding global minimum (or several lowest minima) in configuration space. To adress this problem state-of-the-art CSP methods use evolutionary methods (USPEX \cite{LYAKHOV20131172}, PyXtal\cite{pyxtal}) or particle swarm optimization (CALYPSO\cite{PhysRevB.82.094116}). For energy evaluation and optimization of the candidate structure these methods are originally based Density Functional Theory (DFT).
The main problem with crystal structure prediction (CSP) is associated with the high dimension of the search space. The number of structures to be evaluated while search (and consequently computational costs) grows exponentially with the number of atoms and chemical species \cite{lyakhov_crystal_2010}. That is why reliable crystal structure prediction is limited by relatively small structures with usually two or three chemical components.       

Probably the most popular approach to accelerate CSP is associated with use of machine-learning models of interatomic interaction \cite{mlpots,mlpots1}.
So far, this approach has been applied to predict a dozen of new materials, including boron allotropes \cite{podrogan}, multicomponent and high-entropy alloys \cite{GUBAEV2019148,uspex_ce,srovconi,al-bcc-alloy,kostiuchenko2019-hea}. 

Despite these advances, high-throughput prediction of materials with multiple principal components is still limited due to a huge dimension of the underlying configurational space. The state-of-art methods have been efficient in discovering binary and a number of ternary compounds, but no information on quaternary and other multicomponent structures has been reported to date \cite{uspex_curr}. Recently discovered examples of multicomponent materials, in particular multi-principal element alloys and high-entropy alloys, are believed to possess enhanced refractory and mechanical properties \cite{he1, he2, he3, he4}. Therefore, their discovery is crucial for the technological development.

In this work we limit this formidable problem to discover materials with a (pre-)defined crystal lattice. In this case, the search space narrows down to a set of structures whose atomic positions are near the corresponding lattice sites, and the task consists in discover lattice decorations (i.e., finding out how to assign atomic types to the lattice sites) corresponding to the stable compounds.
The pioneering work in this direction was done by G.W. Hart \textit{et al} \cite{hart1,hart2,hart3}. The proposed an algorithm of generating an exhaustive set of lattice decorations on a set of symmetrically unique supercells, constructed on a fixed crystal lattice.
However, the method was used with DFT calculations which severely limited the configurational space that could be covered.
Later the method was generalized to work with the machine-learning potentials \cite{GUBAEV2019148} that accelerated the search by a factor of 100 to 1000, which, however, was still limited to a few hundred thousand structures which for the case of quartenary compounds corresponds to small unit cells, with at most nine atoms.

In this work we propose a new crystal structure prediction method, which we call CSP-on-lattice. The method is based on a new structure generation algorithm and on-lattice machine learning interatomic potentials for evaluating configurational energy. Unlike the algorithm proposed by G.W. Hart \textit{et al}, our algorithm generates a structure iteratively, atom-by-atom, starting from a single atom. In this sense, the process of structure generation can be called ``structure growth'', which is performed on a fixed lattice. On each iteration of structure growth, the energy is evaluated using cluster expansion or low-rank potential and the optimal stoichiometry is chosen automatically (in other words, it is not a fixed-composition search). 
To alleviate the effect of the spurious free surface occurring at the boundary of the growing structure, we introduce effective interaction with the ``averaged'' species which we call alchemical interaction, following the idea of alchemical energy derivatives that was introduced recently and successfully applied to prediction of energies of small molecules \cite{alc1,alc2}.
This makes our algorithm to resemble the recently popular diffusion algorithms for image generation \cite{diff1, diff2}.
Also, with the help of machine-learning potentials we guide the structure being generated towards the energy minimum, similarly to how the image is guided using the scoring function of the image diffusion algorithms.
 
During the structure growth, as identical atomic neighborhoods appear, we form the supercell cell vectors by connecting such neighborhoods, until we have all three supercell vectors (they are in fact the unit cell vectors of the structure we are growing but we will use the term ``supercell'' not to confuse it with the unit cell of the underlying, fixed lattice that we are decorating).
The proposed method does not require any crystal structure prototypes and the input variables are chemical elements and type of crystal lattice.

The proposed approach was tested on Nb-Mo-Ta-W system. Within this chemical space, we predicted binary, ternary, and quaternary alloys of different stoichiometries and various supercells.
The performance of the algorithm is evaluated by comparing our results to those obtained with high-throughput DFT calculations from the AFLOW database. 
Along with the full four-component system considered the Nb-W and Mo-Ta-W subsystems, and we have discovered new binary, ternary and quaternary structures.

\section{Results and Discussion}

To train CE and LRP we post-relaxed the generated structures with DFT.
To compute reference energies for the training, VASP 5.4.4 \cite{VASP1, VASP2, VASP3} was used.
In our calculations the projector augmented wave (PAW) \cite{paw} method utilizing the Perdew-Burke-Ernzerhof generalized gradient approximation (PBE-GGA) \cite{gga} was employed. The value of plane-wave cutoff energy was set to 400 eV, which is 1.8 times larger than the highest ENMAX energy of the utilized PAW pseudopotentials. To automate calculations for different cell shapes and sizes, we employed an automatic generation of k-mesh by setting \texttt{KSPACING} to 0.13. Both ionic and cell relaxations were included. 
The energy convergence criteria for these types of relaxations was set to $10^{-5}$ eV. For computational efficiency all the selected configurations had no more than 16 atoms in the unit cell. 

For each chemical system the initial lattice parameter was chosen as a mean of lattice parameters of unary structures. For example, taking $a_{\text{Nb}} = 3.30$ \AA\  and $a_{\text{W}} = 3.16$ \AA, the initial lattice parameter for Nb-W alloys is $a_{\text{Nb-W}} = 3.23$ \AA. 

\subsection{Nb-W}

\begin{figure*}[t]
 \includegraphics[width=0.7\textwidth]{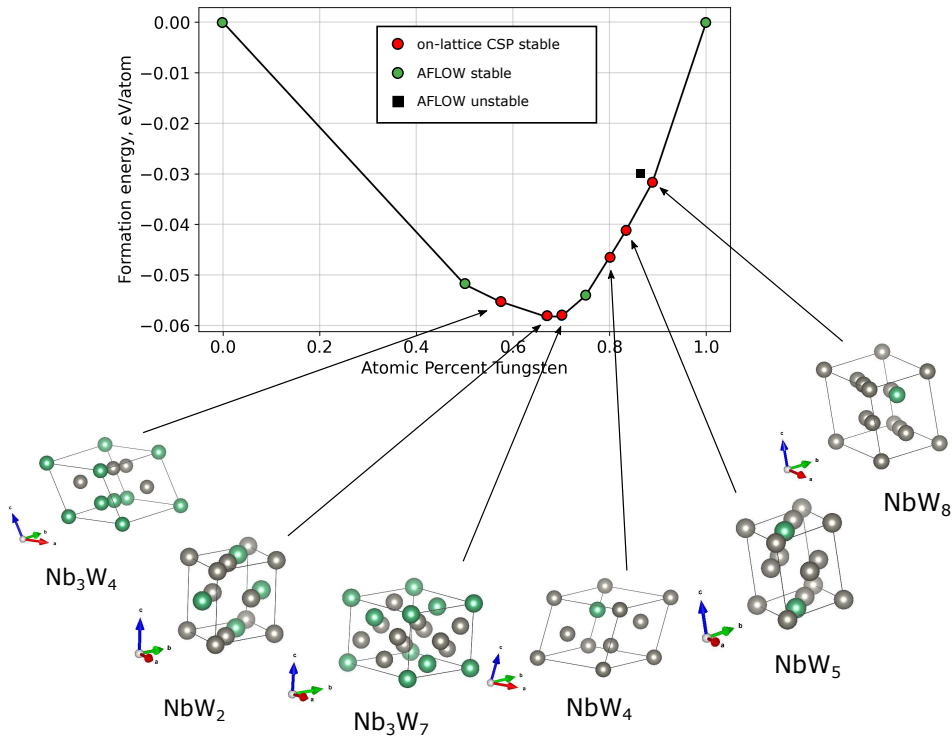}
 \caption{Final DFT Nb-W convex hull and unit cells of new structures. We have discovered six new structures with on-lattice CSP (marked with red circles).}
 \label{fig:fig_nbw}
 \end{figure*}
 
We first apply our algorithm to the construction of the Nb-W convex hull. According to the AFLOW database, this system has three stable phases against unaries, namely, NbW, NbW$_{3}$ and NbW$_{7}$. Both pure Nb and W, as well as the stable binaries have the underlying bcc lattice, so this system is a good test of whether our method is able to predict the stable structures and reproduce the AFLOW convex hull.

 \begin{table}[b]
 \begin{center}
\begin{tabular}{|c|c|} 
\hline
Composition & Position below convex hull, meV/atom\\
\hline
Nb$_3$W$_4$ & $-2.7$\\ 
NbW$_2$ & $-4.8$\\ 
Nb$_3$W$_7$ & $-4.6$\\
NbW$_4$ & $-1.3$\\
NbW$_5$ & $-2.0$\\
NbW$_8$ & $-3.4$\\
\hline
\end{tabular}
\caption{Position below convex hull of predicted Nb-W alloys.}
\label{tab:nbw}
\end{center}
\end{table}

The initial training set consisted of 3 configurations---unary Nb and W with one atom in the unit cell and the binary B2(Nb;W) structure---we deliberately chose it because it is the B32(Nb;W) structure that is on the convex hull.
During the simulation we sampled additional 114 configuration from LRP/CE convex hull. Thus, the final training set consisted of 110 configuration raising a RMSE of 5.31 meV/atom.

Fig. \ref{fig:fig_nbw} shows the final DFT convex hull. As we see, the on-lattice CSP was able to predict all the stable phases reported in AFLOW (green and black markers), as well as six new structures (red markers).
Among the new compositions we discovered Nb$_{3}$W$_{4}$, NbW$_{2}$, Nb$_{3}$W$_{7}$, NbW$_{4}$, NbW$_{5}$ and NbW$_{8}$. Their distance from AFLOW convex hull, measured in meV/atom, is reported in Table \ref{tab:nbw}.  As we see, the new stable compositions were found within a W-rich area of a phase diagram. However, the ground states with the largest distance from AFLOW convex hull, such as Nb$_{3}$W$_{4}$, NbW$_{2}$, Nb$_{3}$W$_{7}$, are observed closer to the center of a phase diagram, which indicates a significant contribution of Nb additions to a phase stability. The ground states discovered in the W-rich region, namely, NbW$_{4}$, NbW$_{5}$ and NbW$_{8}$, are quite shallow with the average distance of 2.2 eV/atom below AFLOW convex hull.
Moreover, we observed that the NbW$_7$ ground state, reported in AFLOW database, became unstable having the distance of 2.3 meV/atom above.\\
Also, we note that we conducted an additional test in which we remove from the training set the NbW and NbW$_3$ structures as present in the AFLOW convex hull, train a new potential and perform two runs of our algorithm.
Interestingly, the algorithm preduced exactly the two removed structures in these two runs, which demonstrates that the algorithm is indeed able to generate the most promising structures first.

\subsection{Mo-Ta-W}

 \begin{figure*}[t]
 \includegraphics[width=0.7\textwidth]{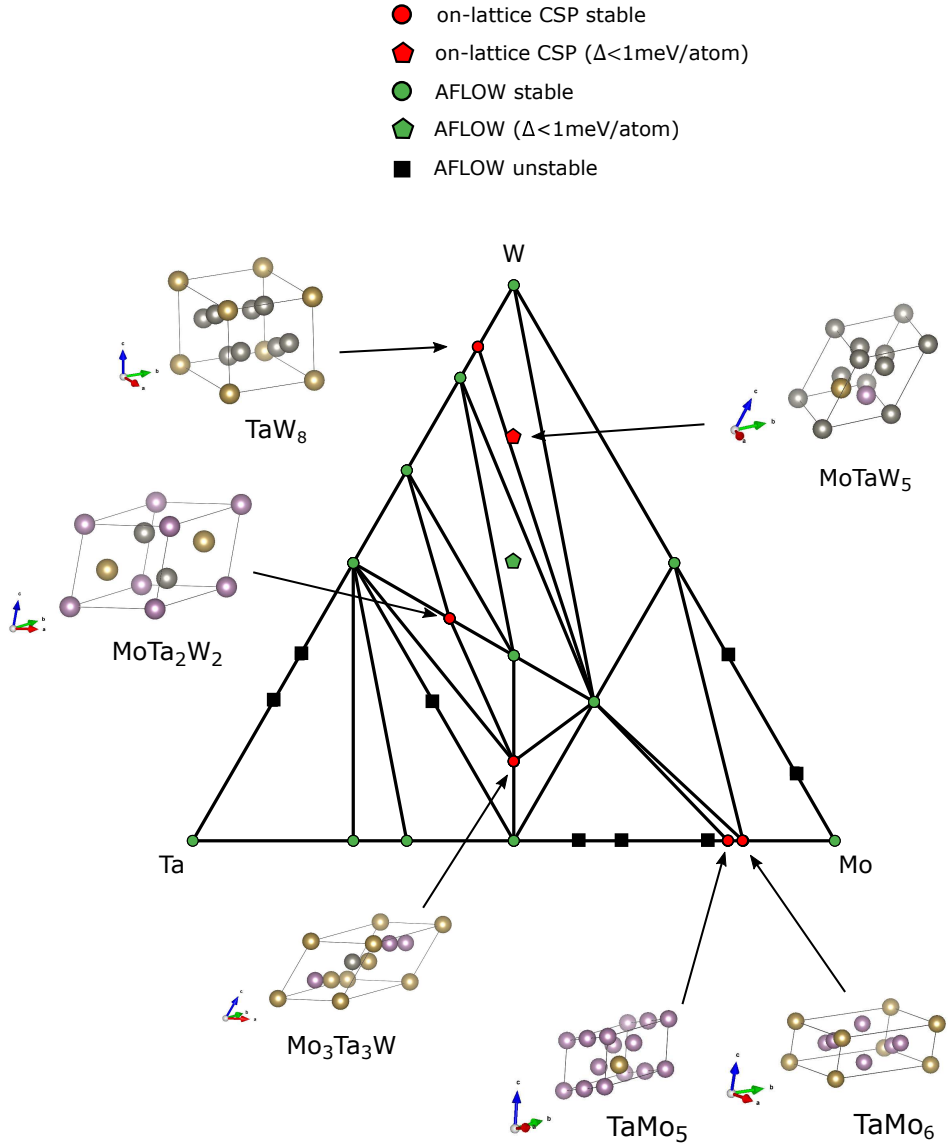}
 \caption{Final DFT Mo-Ta-W convex hull and new structures. We have discovered six new structures whose unit cells are shown (marked with red markers on the convex hull).}
 \label{fig:fig_motaw}
 \end{figure*}

We next test our algorithm on constructing the convex hull for the ternary Mo-Ta-W system. We first constructed the convex hull for the three binary subsystems, namely Mo-Ta, Mo-W and W-Ta and after that performed the simulations for Mo-Ta-W. The initial training set for Mo-Ta-W contained 4 configurations - unary Mo, Ta and W with 1 atom in the unit cell and a random MoTaW with 3 atoms in the unit cell. During the simulations 109 additional configurations were selected, including both binary and ternary structures in nearly equal proportion. The RMSE on the final training set was 8.34 meV/atom.

\begin{table}[b]
\begin{center}
\begin{tabular}{|c|c|} 
\hline
Composition & Formation enthalpy, meV/atom\\
\hline
MoTa & $-187$\\ 
Mo$_3$Ta$_3$W & $-178$\\ 
MoTaW & $-145$\\
MoTaW$_2$ & $-114$\\
MoTaW$_5$ & $-69$\\
\hline
\end{tabular}
\caption{Formation enthalpies of MoTa-based alloys.}
\label{tab:tern_form}
\end{center}
\end{table}

The final DFT convex hull is presented on Fig. \ref{fig:fig_motaw}. It has all the structures present in AFLOW as well as two new ternary phases---Mo$_3$Ta$_3$W, MoTa$_2$W$_2$, and three binary---TaMo$_5$, TaMo$_6$, and TaW$_8$. Since DFT was performed with automatic k-mesh generation, we account for a noise in our calculations and, thereby, also add the structures obtained within 1 meV/atom above the convex hull. Among such structures presented on Fig. \ref{fig:fig_motaw}, MoTaW$_5$ was newly discovered by on-lattice CSP, while MoTaW$_2$, an AFLOW ground state structure, became near-stable after the new configurations were added.
Among the new phases, Mo$_3$Ta$_3$W have the lowest formation enthalpy of $-178$ meV/atom. Remarkably, MoTa-rich region of a phase diagram has another low-lying ground state, namely, equimolar MoTa with fomation enthalpy of $-187$ meV/atom. Such findings reveal that MoTa-rich phases might possess better phase stability, than structures from other regions of a phase diagram. This trend can be observed in Table \ref{tab:tern_form}, where the formation enthalpies for the phases with fixed concentration of Mo and Ta are presented. As it is seen, the formation enthalpy increases from $-187$ meV/atom when no W is present in an alloy, to $-69$ meV/atom when concentration of W exceeds 70 \%.

\subsection{Nb-Mo-Ta-W}

\begin{figure*}[t]
 \includegraphics[width=0.7\textwidth]{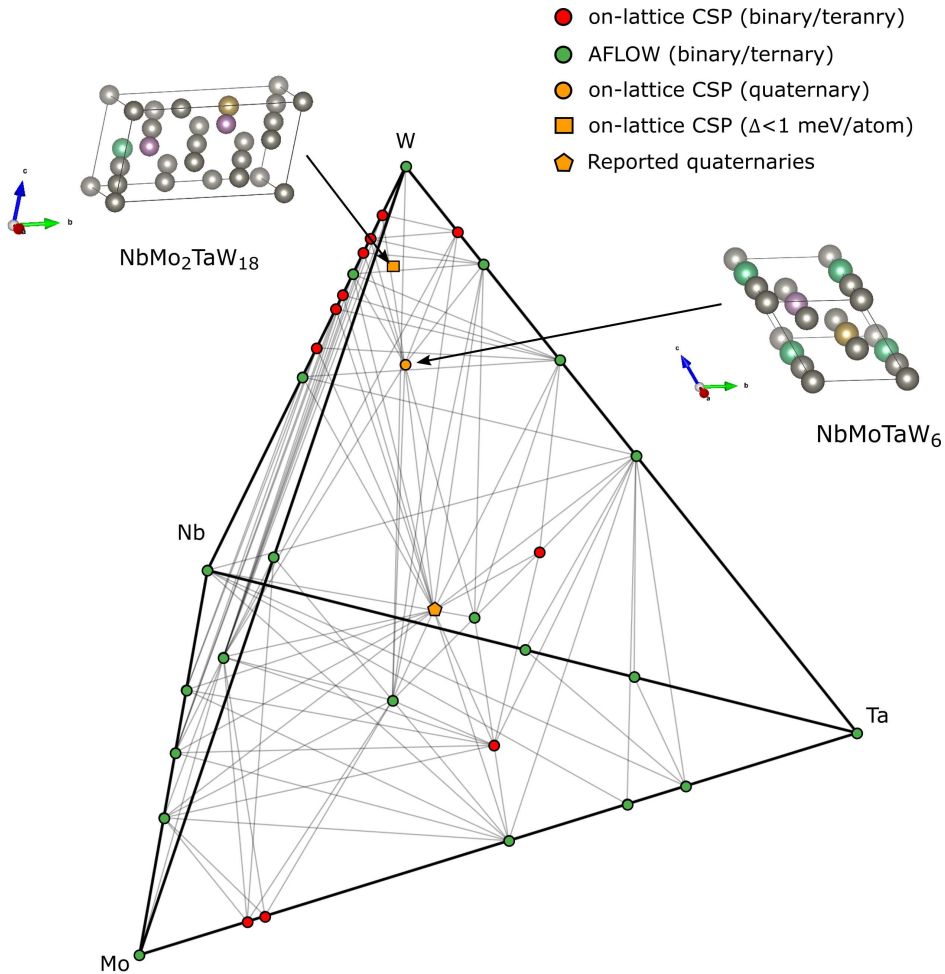}
 \caption{Final DFT Nb-Mo-Ta-W convex hull and unit cells of new structures. We have discovered two new quaternary structures (9-atom NbMoTaW$_6$ and 22-atom NbMo$_2$TaW$_{18}$). Stable structures discovered with on-lattice CSP are marked with orange circles and squares.}
 \label{fig:fig_nbmotaw}
 \end{figure*}

Finally, we applied our method to construct the quaternary Nb-Mo-Ta-W convex hull. For the evaluation of energy of quaternary systems we used LRP as it requires less DFT data for obtaining an adequate approximation accuracy as compared to CE.
We constructed the initial training set in a similar way. That is, it had five configurations---unary Nb, Mo, Ta and W as well as a random quaternary NbMoTaW.
After the simulation the training set was extended to 380 configurations among which 114 structures had quaternary composition. The RMSE on this training set was 7.1 meV/atom.

The final DFT convex hull, post-relaxed with DFT, is presented on Fig. \ref{fig:fig_nbmotaw}. It contains all the structures previously discovered by on-lattice CSP as well as all the stable phases reported in AFLOW. Compositions and corresponding energies from the simplices not mentioned in this work were taken from AFLOW. For a direct comparison all the structures were relaxed using the same DFT settings (pseudopotentials, k-mesh, etc.). Also, we added a quaternary NbMo$_2$Ta$_2$W$_2$ ground state phase, reported by M. Widom \cite{widom1}, which has the formation enthalpy of $-145$ meV/atom. As it is seen we have discovered one new quaternary ground state composition---NbMoTaW$_{6}$ with the formation enthalpy of $-78$ meV/atom. Interstingly, NbMoTaW$_{6}$ as well as the majority of other discovered structures are W-rich phases, which makes their discovery relevant to the application of W-based alloys.

Carefully studying the structures, generated by our algorithm, we observed some amount of configurations with relatively large unit cells ($>$20 atoms). However, all of them were lying higher than 5 meV/atom above LRP convex hull. To investigate their stability, we increased the gap to 10 meV/atom and selected the most low-lying for DFT post-relaxation. Among them, we discovered a 22-atom near-stable NbMoTa$_2$W$_{18}$ with distance above convex hull of 0.9 meV/atom. Such findings indicate, that our algorithm is also capable of discovering stable structures with large unit cells without employing extensive computational resources.

\section{Conclusion}
We have developed a novel generative method for predicting new stable multicomponent alloys and constructing a convex hull, based on a new structure generation algorithm and on-lattice interatomic potentials (CE or LRP). The structure generation is based on iteratively adding atoms on the lattice sites of an empty supercell. By guiding the growth process with energy evaluation by CE or LRP, CSP-on-lattice is able to generate optimal structures, avoiding sampling unpromising regions of configurational space.
The algorithm, to some extent resembles the diffusion generation algorithms for images \cite{diff1,diff2}.
Additionally, it is sufficient for the method to use a relatively small amount of DFT calculations as training data.
We validated our method on constructing the convex hull for Nb-W, Mo-Ta-W, and Nb-Mo-Ta-W systems. In each of them we discovered several new stable structures.
To discover the stable Nb-W and Mo-Ta-W structures we needed less than 200 DFT calculations.
For the Nb-Mo-Ta-W system we performed less than 400 DFT calculations for the LRP training before a new stable quaternary composiions were detected. Such results indicate an ability of our method to perform high-throughput discovery of multicomponent materials using a limited amount of DFT data.

\section{Methodology}

\subsection{On-lattice CSP}

We start with a brief conceptual description of our method, and in the following subsections we present the details.
In our algorithm we ``grow'' the structure, by adding atoms sequentially to an empty configuration in the sites of the lattice (bcc in our case).
Each time a new atom is added, we identify identical atomic neighborhoods and construct the unit cell vectors by connecting the centers of these neighborhoods.
Once all three vectors are constructed, the final crystal structure is obtained. Further in the section the detailed description of the algorithm will be given for a binary system. For the clarity all illustrations will be two-dimensional.

\begin{figure*}[t]
 \includegraphics[width=0.85\textwidth]{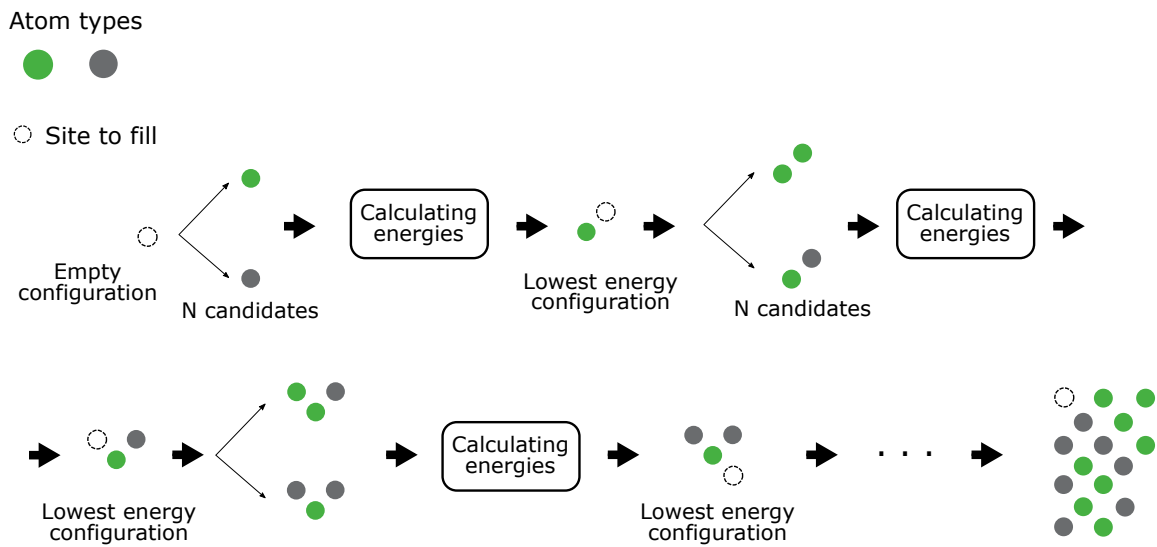}
 \caption{Schematic illustration of structure growth. Atoms of two different chemical types are marked with green and grey colors. Empty lattice site is marked with a dashed circle.}
 \label{fig:growth}
 \end{figure*}

The structure growth procedure is presented on Fig.\ref{fig:growth} and detailed in this section.
The process starts with an empty configuration. Next, an atom of each chemical type is added in an empty lattice site. This produces a population of $N$ candidate structures (two in the binary case). After that, their energies are calculated and the lowest energy structure is selected.
We note that at this stage configuration is not complete and the details of how to evaluate energy of such configurations are given in sections \ref{energy}, \ref{alch}.
Then, an atom of each chemical type is added to the selected structure, producing a new population of $N$ candidates. Their energies are calculated and the lowest energy structure is again selected. Note that each new atom is added in the nearest-neighboring site of the previously added atom.
This process continues until we find two identical atomic neighborhoods and we then construct a supercell vector as described in the next subsection. 

\begin{figure*}[t]
 \includegraphics[width=0.8\textwidth]{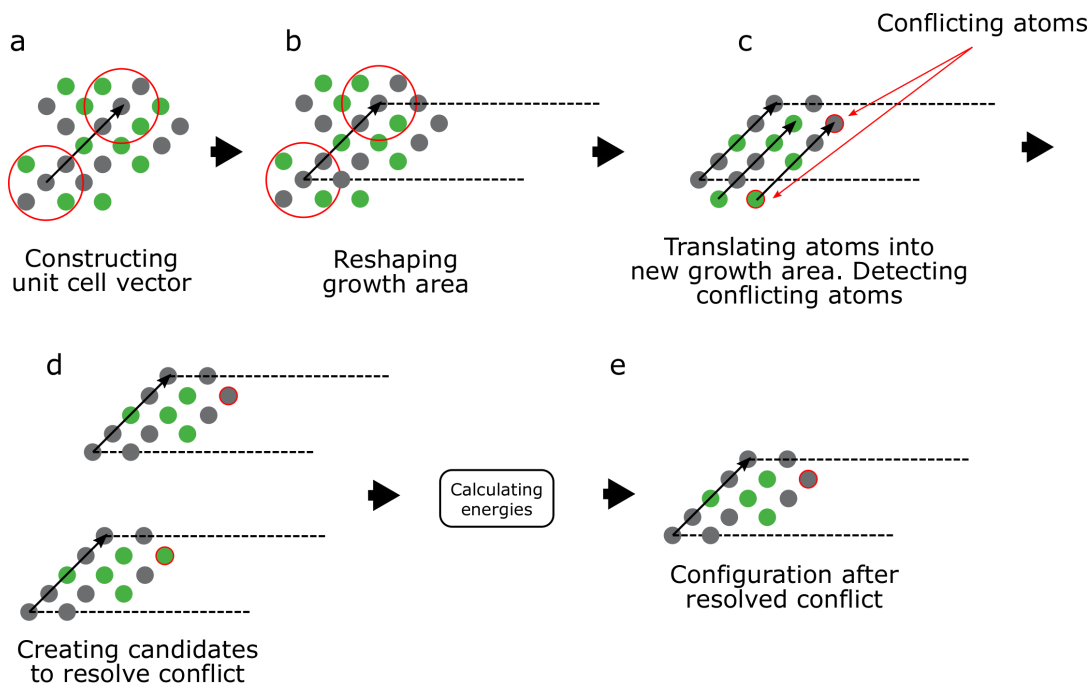}
 \caption{Schematic illustration of the unit cell vector construction process. Atoms of two different chemical types are marked with green and grey colors. Identical atomic neighborhoods are encircled with red.}
 \label{fig:un_cell}
 \end{figure*}

Every time a new atom is added to a configuration we search for two identical atomic neighborhoods. Once they are discovered, a supercell vector is constructed by connecting the centers of these neighborhoods (Fig.\ref{fig:un_cell}a). Next, we reshape the growth area using the constructed vector, as it is shown on \ref{fig:un_cell}b. After, as seen from Fig.\ref{fig:un_cell}c the atoms are translated into the new growth area. During this process ``conflicting'' atoms can appear. These are the atoms of different chemical type that share the same lattice site after translation. In this case we add an atom of the chemical type that results in a lower energy (Fig.\ref{fig:un_cell}d-e).

 \begin{figure*}[ht]
 \includegraphics[width=0.8\textwidth]{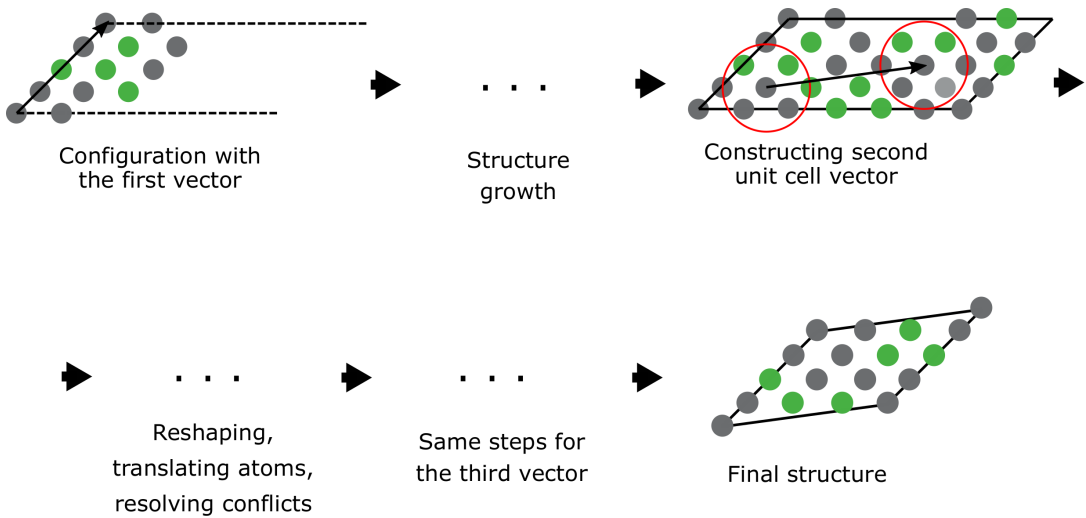}
 \caption{Schematic illustration of constructing the rest of the vectors and obtaining the final structure. Atoms of two different chemical types are marked with green and grey colors. Identical atomic neighborhoods are encircled with red.}
 \label{fig:final}
 \end{figure*}

After the first vector is constructed and all conflicts are resolved, the structure growth continues within a new area. Once a new pair of identical atomic neighborhoods is discovered, the second unit cell vector is constructed. Then we again reshape the growth area, translate atoms and resolve the conflicts. Finally, all the same steps are performed until the third vector (which is needed in three dimensions) is constructed. Once all the three supercell cell vectors are constructed, the algorithm adds atoms in the remaining empty sites (if such are present) which is the last step in obtaining the final structure. The schematic illustration of these steps are presented on Fig.\ref{fig:final}.

\subsection{Evaluation of energy}\label{energy}
In our simulations an atomistic configuration is represented by an ideal crystalline lattice with sites occupied by one of the chosen atomic species. Therefore, we use on-lattice interatomic potentials to evaluate the energy. In these models the total energy is represented as a sum of energy contributions from individual atomic neighborhoods and is given as follows:
\begin{eqnarray}
E = \sum_{\xi \in \Omega} V(\sigma_{1}, ... ,\sigma_{n})
\label{eq:tot_E}
\end{eqnarray}
where $\xi$ is the central atom of the neighborhood; $\Omega$ are lattice sites; $V$ is the tensor of size $m^n$ that stores energy contributions of atomic neighborhoods with $m$ chemical types and $n$ neighbors; atomic neighborhood is denoted as $(\sigma_{1},...,\sigma_{n})$, where $\sigma_{1}, ... ,\sigma_{n}$ are relative atomic types ($0,1,2,...,n$). 
Although in our representation the lattice sites are ideal, in the subsequent DFT calculations that are discussed further, relaxation of an ideal structure is performed.

We use two on-lattice methods to calculate the configurational energy, namely, the cluster expansion (CE) and the low-rank potential (LRP) \cite{SHAPEEV201726}. While CE is applied to calculate the energies of binary and ternary alloys, the LRP is used for the quaternary ones. We motivate this choice by a higher efficiency of the LRP in modeling many-body interactions in multicomponent systems.

In the CE model the energy contribution of an atomic neighborhood is given using the following expression:
\begin{eqnarray}
	V(\sigma_{1}, ... ,\sigma_{n}) = 
	\sum_{ij}J_{ij}\sigma_{i}\sigma_{j}+
	\sum_{ijk}J_{ijk}\sigma_{i}\sigma_{j}\sigma_{k}+ ... , \label{eq:V_CE}
\end{eqnarray}
where,  $J_{i,j}$ and $J_{i,j,k}$ are the so-called effective cluster interactions (ECIs) of pair and triplet atomic clusters of the neighborhood; $\sigma_{i}$, $\sigma_{j}$ and $\sigma_{k}$ are types of atoms $i$, $j$ and $k$. 

	In the LRP the energy contribution of an individual neighborhood is given in a tensor-train format and reads:
	\begin{eqnarray}\label{eq:V_LRP}
		V(\sigma_{1}, ... ,\sigma_{n}) = \prod_{i}A_{i}(\sigma_{i}),
	\end{eqnarray}
where $A_{i}$ are matrices with size $r$ or less, that contain the parameters of the model. The size of $A_{1}$ and $A_{n}$ is $1\times r$ and $r\times1$ respectively, and the sizes of $A_{2},...,A_{n}$ are $r\times r$, so that their product gives a scalar, which corresponds to the energy contribution of a neighborhood. To reduce the number of required DFT data LRP is represented in a tensor-train format which reduces the number of parameters from $m^n$ to only $nmr^2$. Thus, the predictive accuracy of LRP is influenced by two hyperparameters, namely, the decomposition rank $r$ and number of neighbors $n$.
In this work we set $n=9$, which corresponds to coordination number in bcc lattice including the central site. The value of rank $r$, was set to $r=3$, which eventually produced 324 independent parameters in the LRP model.
	
The parameters of CE and LRP are found by minimizing the following functional,  
	\begin{eqnarray}\label{eq:KL}
		\sqrt{\frac{\sum_{k=1}^{K}(E(\sigma^{(k)}) - E^{qm}(\sigma^{(k)}))^2}{K}},
	\end{eqnarray}
where $\sigma^{(k)}$ are the training atomic configurations; $K$ is the training set size; $E^{qm} (\sigma^{(k)})$ is the quantum-mechanical energy; $E(\sigma^{(k)})$ is the energy predicted by CE or LRP. 

\subsection{``Alchemical'' potential}\label{alch}
During the structure growth stage, the added atoms have incomplete atomic neighborhoods.
Such atoms create a substantially large surface which, in turn, cause an increase in the interaction energy (due to a large artificial surface energy).
Due to this the algorithm might add atoms of suboptimal chemical type during structure growth stage.

\begin{figure*}[t!]
 \includegraphics[width=\textwidth]{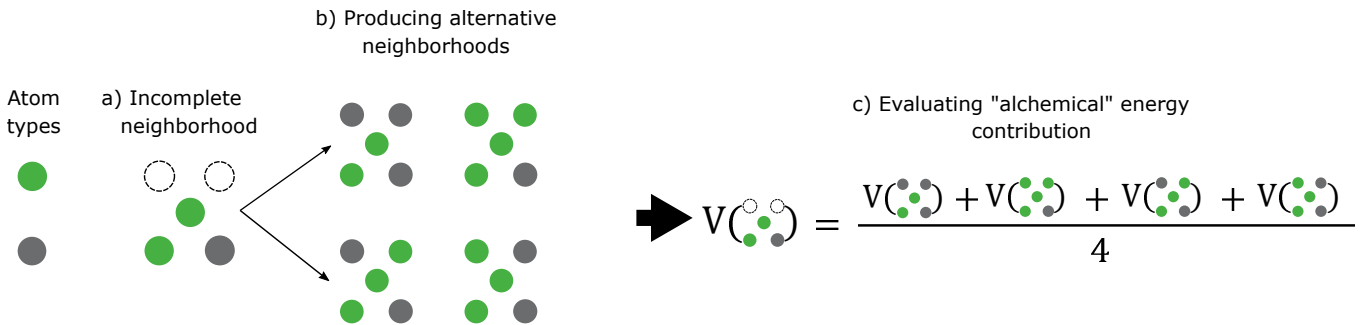}
 \caption{Schematic illustration of evaluation of "alchemical" energy contribution. The example is given for a binary system. a) Detection of an incomplete neighborhood. b) Substitution of atomic species into empty lattice sites. c) Evaluation of "alchemical" energy contribution. Atoms of two different chemical types are marked with green and grey colors. Empty sites are marked with a dashed circle.}
 \label{fig:fig_alc}
 \end{figure*}

To minimize this effect we adopted a strategy which we call the ``alchemical'' potential.
It consists of filling empty sites of an atomic neighborhood with an imaginary (``alchemical'') atom type which is, in a way that we detail below, a superposition of the real chemical types.
Addition of ``alchemical'' atoms in the empty sites of the neighborhood creates the so-called an ``alchemical'' atomic neighborhood, whose energy contribution is calculated as follows:
\begin{equation}
    V(\sigma_{al}) = \frac{V(\sigma_{1}) + ... + V(\sigma_{m^N})}{m^N}
\end{equation}
where $V(\sigma_{1})$, ...  $V(\sigma_{m^N})$ are energies of neighborhoods, in which empty sites are filled with atoms of real chemical types; $m^N$ is the number of such neighborhoods, where $N$ is the number of empty sites in incomplete neighborhood. The process of evaluating the energy of an ``alchemical'' neighborhood is illustrated on Fig.\ref{fig:fig_alc}.

\subsection{Biasing potential}
On-lattice-CSP generates only one configuration during a single run. Hence, if the simulation is restarted with the same potential, the algorithm will generate the same configuration each time. We hence bias the LRP/CE potential after each run of the algorithm in order to ensure the prediction of different structures. The biasing consists of adding a constant value $\delta$ to the energy contribution of each atomic neighborhood present in the predicted structure.
Therefore, during the next run the previously predicted structure will have a higher energy and the algorithm may adopt a different structure generation path (if it happens that the same path is chosen, the resulting structure will be biased again until the biasing will result in the new structure).
In our simulations the value of $\delta$ was set to 0.001 eV.

\begin{figure*}[t!]
\center
 \includegraphics[width=0.5\textwidth, height=12cm]{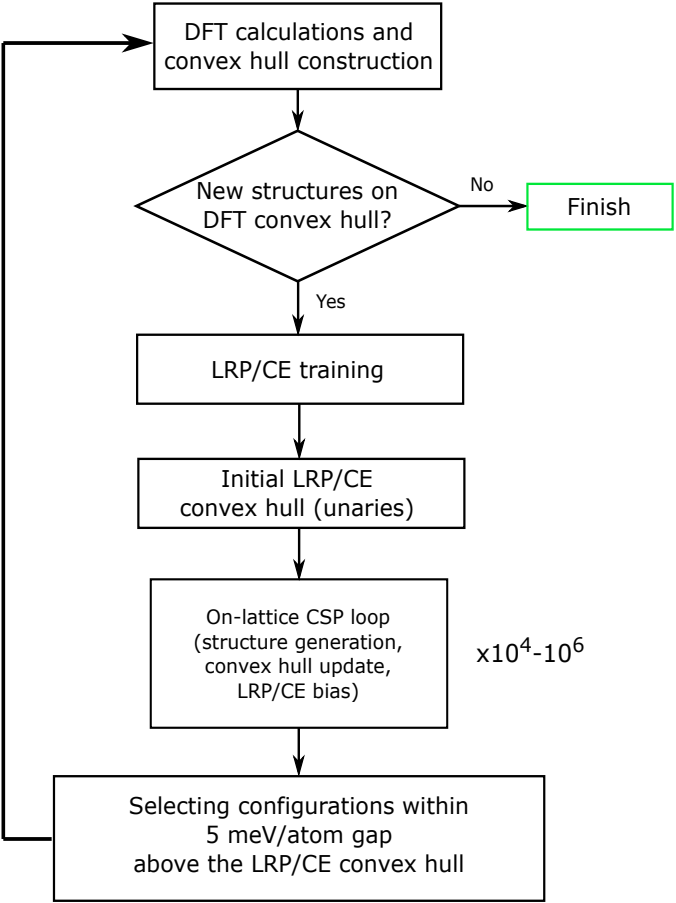}
 \caption{Convex hull construction algorithm.}
 \label{fig:ch_construction}
 \end{figure*}

 \begin{figure*}[t]
 \includegraphics[width=0.8\textwidth]{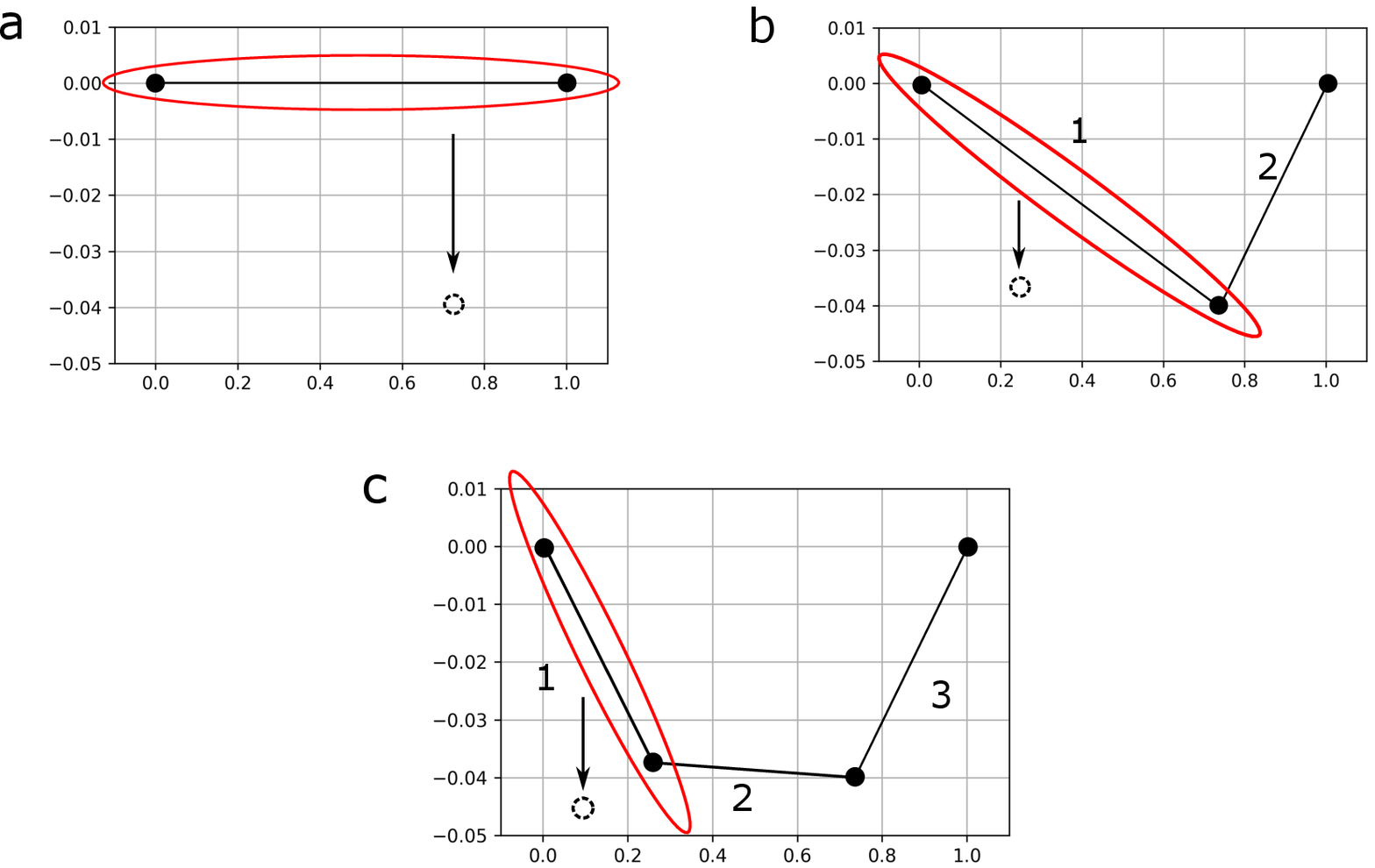}
 \caption{Convex hull update during on-lattice CSP simulation. Current simplex is encircled with red. A potentially new point on convex hull is marked with a dashed circle.}
 \label{fig:ch_update}
\end{figure*}

\subsection{Convex hull construction}
The convex hull construction algorithm is presented on Fig.\ref{fig:ch_construction}.
We start with the initial DFT convex hull and then we construct the LRP/CE convex hull with on-lattice CSP as explained in the next paragraph.
After that, we select all the structures within 5 meV/atom interval, post-relax them with DFT, and add to the initial DFT convex hull.
The simulation is running until the DFT convex hull is updating.

The details of constructing the LRP/CE convex hull are as follows.
On-lattice CSP simulation starts with the initial LRP/CE convex hull and iteratively updates each simplex. The schematic illustration of this process is presented on Fig.\ref{fig:ch_update}. As seen on Fig.\ref{fig:ch_update}a, we choose a simplex of the initial convex hull and try to update it by generating a configuration with the energy below this simplex (Fig.\ref{fig:ch_update}a). After, in the similar manner, we try to update the simplex 1 of the new convex hull (Fig.\ref{fig:ch_update}b-c) and so on. This way we iterate over each simplex. On average it takes about $10^4$--$10^6$ iterations to construct the final LRP/CE convex hull.

Every time a simplex is chosen, we calculate the energy of its vertices, and set up the chemical potential of each species subtract them from tensor $V$ (thus simply introducing a constant species-dependent shift in the energy, see Section \ref{energy}), so that the energy of each end point of the simplex is zero.

\section{Funding}\label{sec:funding}

This work was supported by Russian Science Foundation (grant number 23-13-00332, \url{https://rscf.ru/project/23-13-00332/}).

\section{Competing Interests}
The Authors declare no Competing Financial or Non-Financial Interests.

\section{Availability of Data and Material}
The access to the data and materials used for reproduction of results can be granted upon request.
	
\section{Code Availability}
The access to the software required for the reproduction of results can be granted upon request.
\\
\section{Author Contributions}
V.S., E.V.P and A.V.S. developed the code for on-lattice CSP. V.S. performed the simulations and analyzed the results. All authors participated in the preparation of the manuscript.

\bibliography{main}
\end{document}